\documentclass[a4paper]{article}

\usepackage{INTERSPEECH2021}
\usepackage{graphicx}
\usepackage{blindtext}
\usepackage{array}
\usepackage{multirow}
\usepackage{subcaption}
\usepackage{threeparttable}
\graphicspath{ {./} }

\title{Streaming Transformer for Hardware Efficient Voice Trigger Detection and False Trigger Mitigation}
\name{Vineet Garg$^{\ast}$, Wonil Chang$^{\ast}$, Siddharth Sigtia, Saurabh Adya, Pramod Simha, Pranay Dighe, Chandra Dhir}

\address{
  Apple
}
\email{\{vineetgarg, wonil\_chang, sidsigtia, sadya, psimha, pdighe, cdhir\}@apple.com}

\begin{document}

\maketitle
\renewcommand{\thefootnote}{\fnsymbol{footnote}}
\footnotetext[1]{Equal contribution.}
\begin{abstract}

We present a unified and hardware efficient architecture for two stage voice trigger detection (VTD) and false trigger mitigation (FTM) tasks. Two stage VTD systems of voice assistants can get falsely activated to audio segments acoustically similar to the trigger phrase of interest. FTM systems cancel such activations by using post trigger audio context. Traditional FTM systems rely on automatic speech recognition lattices which are computationally expensive to obtain on device.
We propose a streaming transformer (TF) encoder architecture, which progressively processes incoming audio chunks and maintains audio context to perform both VTD and FTM tasks using only acoustic features. The proposed joint model yields an average 18\% relative reduction in false reject rate (FRR) for the VTD task at a given false alarm rate. Moreover, our model suppresses 95\% of the false triggers with an additional one second of post-trigger audio. Finally, on-device measurements show 32\% reduction in runtime memory and 56\% reduction in inference time compared to non-streaming version of the model.

\end{abstract}
\noindent\textbf{Index Terms}: Keyword Spotting, Speech Recognition, Acous- tic Modeling, Neural Networks, Deep Learning, False Trigger Mitigation

\section{Introduction}
\label{sec:intro_new}

Intelligent voice assistants (VA) are becoming a ubiquitous part of our daily interaction with smart phones, smart speakers, and other AI-enabled devices. A VA query starts with a trigger phrase followed by the user’s request.
Many of the on-device VTD systems follow a two stage cascaded design \cite{gruenstein2017cascade}\cite{wu2018monophone}\cite{Sigtia_2020}\cite{adya2020hybrid} to process the request. 
The first stage has a low footprint always-on model \cite{shrivastava2020optimize}\cite{sainath2015convolutional}\cite{sigtia2018efficient}, which processes continuous audio streams to identify candidate audio segments containing the trigger phrase.
Some of these segments can be false triggers that are acoustically similar to the trigger phrase. Since unintended activation of a VA can invade user privacy \cite{2020unacceptable} and degrade user experience, we use a larger capacity model  \cite{Sigtia_2020}\cite{adya2020hybrid}\cite{Kumar2020BuildingAR} as a second stage VTD to decide whether to activate the VA or not. 

Unlike VTD, FTM systems focus on the post-trigger audio to reduce unintended activation of VAs. It has been shown that post-trigger audio has discriminating information to determine the user intent which can be used to achieve higher accuracy decisions for VTD \cite{WangKRKV20}\cite{dighe2020knowledge}\cite{sigtia2020progressive}.
Towards this end, various FTM systems combine acoustic features with automatic speech recognition (ASR) based features 
\cite{mallidi2018devicedirected}\cite{8461478}\cite{Jeon_2019}\cite{agarwal2020complementary}\cite{dighe2020latticebased}\cite{Gillespie_2020} to determine if the user query is intended for the VA or not. Generally, full-fledged ASR systems are available on a server only because they are computationally expensive to be run locally. This raises privacy concerns since unintended queries can leave the user's device.  
On-device FTM systems use only acoustic features \cite{dighe2020knowledge}\cite{norouzian2019exploring}\cite{Tong2020StreamingRW} for hardware efficiency and to avoid such privacy concerns. 





Both VTD and FTM tasks aim to reduce the unintended activation of VAs using two different approaches. 
While VTD has access to the trigger segment, FTM has access to the trigger segment followed by the post-trigger audio. 
To attain the common goal, we propose a joint streaming TF encoder architecture, which progressively processes incoming audio chunks while maintaining the audio context to perform both VTD and FTM tasks simultaneously. Our proposed model uses only acoustic features for both the tasks and removes dependency on server-side ASR. Morevover, our model shares the compute between VTD and FTM tasks leading to reduced run-time memory and latency. This enables the use of our proposed model on low power edge devices without server-side support.  To our knowledge, this work is the first to simultaneously handle the two target tasks on-device in a streaming fashion.

In our experiments, we show that the joint model improves FRR by an average 18\% relative compared to the baseline at the same false alarm operating point for the VTD task. The model also correctly rejects up to 95\% of unintended false triggers with an additional one second of post-trigger audio for the FTM task. The streaming nature of the model allows reuse of audio context, which reduces the run-time memory by 32\% and the inference time by 56\% when processing an additional one second of post-trigger audio. Furthermore, the run-time memory and inference time increase linearly with respect to the audio length, whereas they grow quadratically for the non-streaming version of the model.

The rest of the paper is organized as follows. Section \ref{sec:architecture} provides details of our proposed architecture. Section \ref{sec:training} has the details on training methodology. Section \ref{sec:results} discusses evaluation results and ablation study followed by the conclusions in Section 5.

\section{Model Architectures}
\label{sec:architecture}

\begin{figure}
\centering
\includegraphics[width=\linewidth]{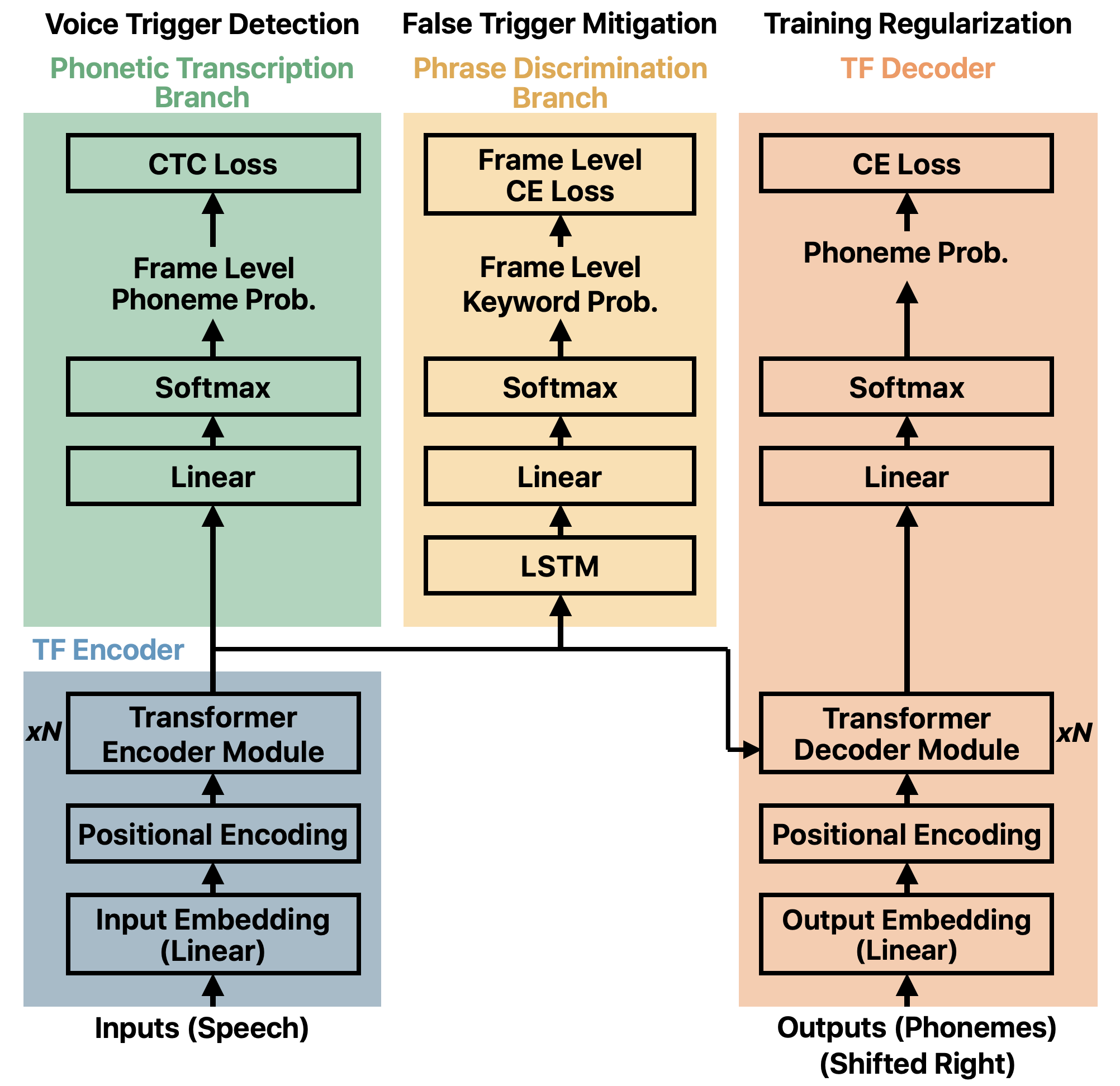}
\caption{Proposed network architecture. Blue block is the backbone of TF encoder. Orange block is an optional auto-regressive TF decoder that can be added during training. Green block is the phonetic transcription branch and is trained by minimizing CTC loss. Yellow block is the phrase discrimination branch and is trained by minimizing frame-level CE loss.}
\vspace{-5mm}
\label{fig:model_arch}
\end{figure}

The outline of the proposed system design with the newly introduced streaming model architecture is depicted in Figure \ref{fig:model_arch}. We focus on optimizing the two targets tasks that are interconnected via the backbone TF encoder module (the blue block in Figure \ref{fig:model_arch}). The backbone extracts acoustic embeddings, which are used by the phonetic transcription branch for the VTD task and phrase discrimination branch for the FTM task. The third branch is an auto-regressive TF decoder which is used only during training as a regularizer. We first describe details of the baseline TF encoder architecture \cite{adya2020hybrid}.

\subsection{TF Encoder: Baseline Architecture}
\label{sec:baseline_arch_tf_encoder}

We use the TF encoder \cite{adya2020hybrid} as baseline, which is a Transformer \cite{vaswani2017attention} model trained with the connectionist temporal classification (CTC) loss on the encoder and cross-entropy (CE) loss on the decoder. We use only the encoder during inference. Since self attention (SA) layers of the encoder look at the entire input context at once, we refer to them as vanilla SA layers in rest of the paper. The trained TF encoder is used to initialize the multi-task learning (MTL) framework, where we add a parallel branch on top of the encoder for phrase discrimination task \cite{Sigtia_2020}\cite{adya2020hybrid}\cite{sigtia2020progressive}. Both the phonetic transcription and phrase discrimination tasks are trained to minimize the CTC loss in their respective branches.

\subsection{Streaming TF Encoder: Proposed Architecture}
\label{sec:proposed_arch}
We build upon the existing TF encoder proposed in \cite{adya2020hybrid} and introduce a novel streaming TF encoder that:
(1) replaces vanilla SA layers with streaming SA layers, (2) introduces frame-wise CE loss in the phrase discrimination branch rather than the CTC loss, and (3) introduces a unidirectional long-short-term memory (uniLSTM) layer in the phrase discrimination branch.

\subsubsection{Streaming SA Layers}
\label{sec:streaming_sa}
\begin{figure}
\centering
  \begin{subfigure}[b]{0.60\linewidth}
    \includegraphics[width=\linewidth]{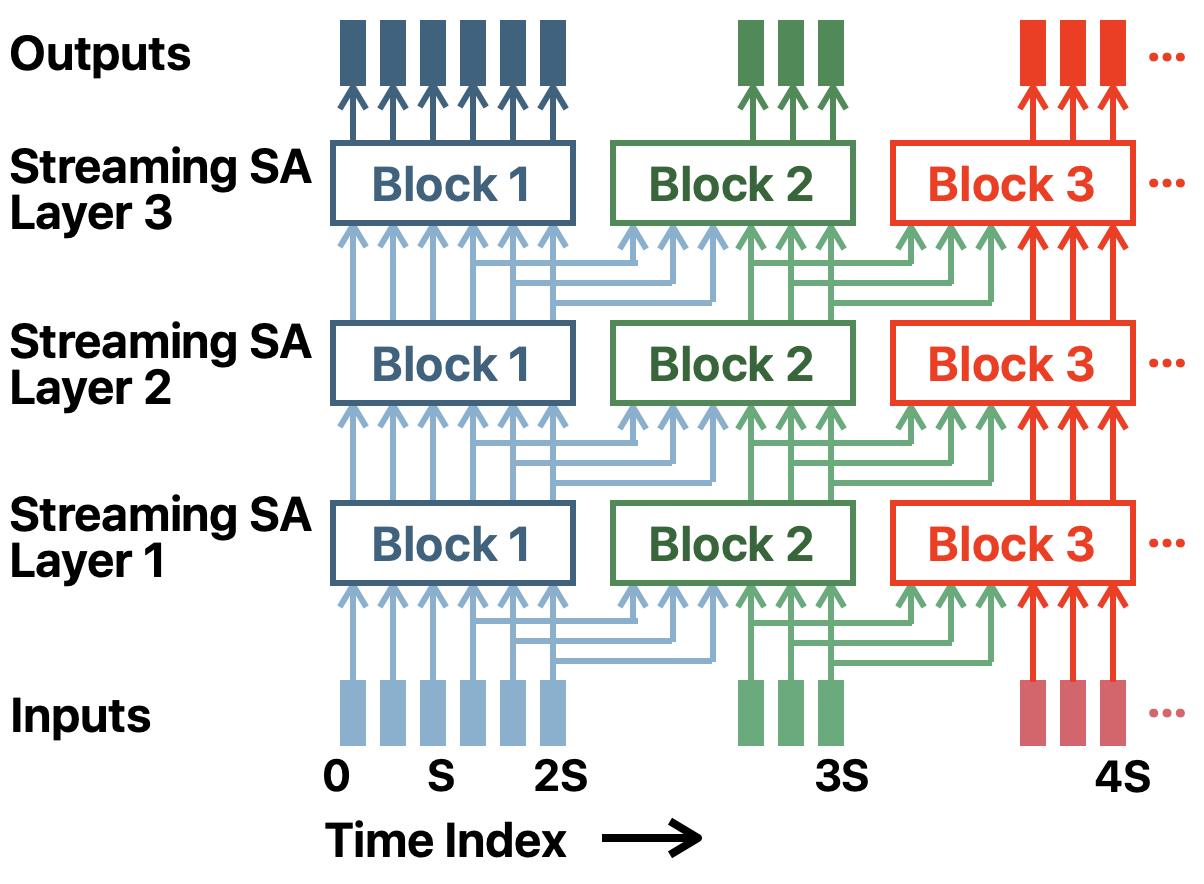}
    \caption{}
    \label{fig:streaming_SA}
  \end{subfigure}%
\hfill
  \begin{subfigure}[b]{0.39\linewidth}
    \includegraphics[width=\linewidth]{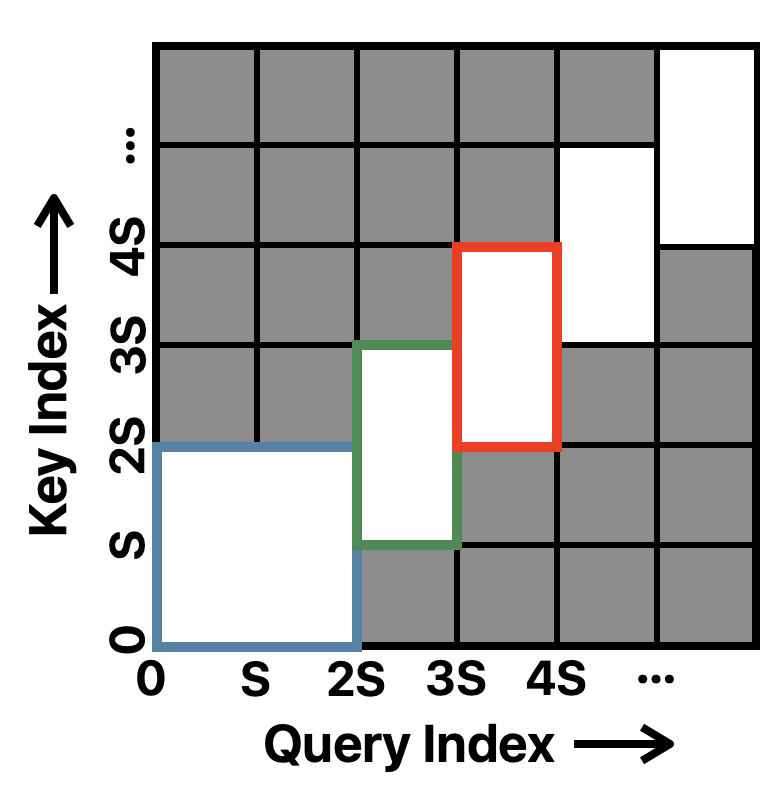}
    \caption{}
    \label{fig:attention_mask}
  \end{subfigure}%

\caption{(a) Block diagram of streaming SA layers. (b) Attention mask to simulate a streaming SA layer. The white regions propagate attention weights. Blue, green, and red rectangle represents the attention weights for Block 1, 2, and 3, respectively. \\ In both (a) and (b), we assume that the block shift $S$ is 50\% of the block size $B$, i.e., $B = 2S$. }
\vspace{-5mm}
\label{fig:streaming_SA_all}
\end{figure}

To enable training and inference of the proposed streaming TF encoder, we replace the vanilla SA layers with streaming SA layers. We follow the block processing protocol presented in \cite{dai2019transformerxl}. The streaming SA layers process the incoming audio in a blockwise manner with certain shared left context, as shown in Figure~\ref{fig:streaming_SA}. 
Let $\mathbf{Q} = \left[ \mathbf{q}_1, \mathbf{q}_2, \ldots  \right]$, $\mathbf{K} = \left[ \mathbf{k}_1, \mathbf{k}_2, \ldots  \right]$, and $\mathbf{V} = \left[ \mathbf{v}_1, \mathbf{v}_2, \ldots  \right]$ represent the query, key, and value matrices for a vanilla SA layer, respectively. 
A streaming SA layer uses $\mathbf{Q}_{i}$, $\mathbf{K}_{i}$, and $\mathbf{V}_{i}$ for the $i^{th}$ block processing. 
We assume that the block shift $S$ is 50\% of the block size $B$, i.e., $B = 2S$.
In the first block, $\mathbf{Q}_{1} = \left[\mathbf{q}_1, \ldots, \mathbf{q}_{2S} \right]$, $\mathbf{K}_{1} = \left[\mathbf{k}_1, \ldots, \mathbf{k}_{2S} \right]$, and $\mathbf{V}_{1} = \left[\mathbf{v}_1, \ldots, \mathbf{v}_{2S} \right]$. For $i \geq 2$, $\mathbf{Q}_{i} = \left[\mathbf{q}_{iS+1}, \ldots, \mathbf{q}_{(i+1)S} \right]$, $\mathbf{K}_{i} = \left[\mathbf{k}_{(i-1)S+1}, \ldots, \mathbf{k}_{(i+1)S} \right]$, and $\mathbf{V}_{i} = \left[\mathbf{v}_{(i-1)S+1}, \ldots, \mathbf{v}_{(i+1)S} \right]$. 
Note that we use last $ S $ vectors of query, last $B$ vectors of key, and last $B$ vectors of value vectors from the second block processing. While $\mathbf{Q}_{i}$ uses the query vectors from the latest block shift, $\mathbf{K}_{i}$ and $\mathbf{V}_{i}$ use the key and value vectors from the previous block shift as well.

We simulate the streaming block processing in a single pass while training by assigning an attention mask (Figure~\ref{fig:attention_mask}) to attention weight matrix of the vanilla SA layer as in \cite{tsunoo2019transformer}. The mask propagates only the white regions of attention weight matrix and generates the equivalent attention output of a streaming SA described above. This helps avoid slowdown of the training by iterative block processing. 
The decoder still has vanilla SA and cross attention layers since the decoder is not used during inference.
While many of the streaming SA models use relative position encoding \cite{dai2019transformerxl}\cite{zhang2020transformer}\cite{wang2021wake}, we use an absolute position encoding scheme from the original TF \cite{vaswani2017attention}.

\subsubsection{CE Loss in Phrase Discrimination Branch}
\label{sec:ce_loss_disc_branch}
In the baseline architecture, we used the sequence to label CTC loss for phrase discrimination \cite{Sigtia_2020}\cite{adya2020hybrid}\cite{sigtia2020progressive}. 
The CTC loss is prone to making decisions in a greedy manner by producing highly peaky and overconfident distributions \cite{NEURIPS2018_e44fea3b}. This can cause streaming SA layers trained with CTC loss (in the phrase discrimination branch) to misclassify the false trigger utterances which are acoustically similar to the trigger phrase as true triggers. 
Post-trigger acoustic context, which can help with accurate phrase discrimination, cannot be used since there is no future audio look-ahead mechanism. This presents an inherent limitation for streaming inference of the phrase discrimination task when the CTC loss is used for training. Therefore, we replace the CTC loss in the phrase discrimination branch with the CE loss. Since the target would have a single label (true or false trigger), we replicate the target labels across all frames and compute frame-wise CE loss.

\subsubsection{UniLSTM in Phrase Discrimination Branch}
\label{sec:lstm_disc_branch}
We introduce a uniLSTM \cite{hochreiter1997long} layer between the TF encoder embeddings and the dense layer in the phrase discrimination branch. UniLSTM has two benefits here; it smoothens the fluctuation of the phrase discrimination output by accumulating the activity of TF encoder embeddings, and it enables long-range context propagation across distant streaming blocks.

\section{Training Methodology}
\label{sec:training}

The training methodology is similar to our baseline TF encoder \cite{Sigtia_2020}\cite{adya2020hybrid}\cite{sigtia2020progressive}. In the first phase, we train a Transformer network where the encoder has streaming SA layers in a sequence-to-sequence fashion for a phonetic transcription task with the CTC loss on the encoder and CE loss on the decoder (blue, green and orange blocks in Figure \ref{fig:model_arch}). The training data has 2700 hours of clean audio with transcribed phonetic labels (54 classes). The audio data is augmented with room impulse responses and echo residuals making a total of 8100 hours of data. We compute 40 dimensional mel-filterbank features at a rate of 100 frames per second. At every frame, we stack the current frame with three frames each for previous and future to create 280 dimensional features. The sequence is sub-sampled by a factor of three and fed to the network. The output of the network is 54 dimensional corresponding to 54 context independent (CI) phones, including start of sequence, end of sequence, word boundary and a blank label. The network has six SA layers with 256 hidden units each and 4 heads with each block followed by 1024 hidden units of a feed forward block. In streaming SA layers, we use a block size $B=64$ (1.92s) frames and move in steps of $S=32$ frames with an overlap of previous $B-S=32$ frames.

In the second phase of training, we remove the TF decoder (orange block in Figure \ref{fig:model_arch}) and add a parallel phrase discrimination branch (yellow block in Figure \ref{fig:model_arch}) on top of the encoder. The new branch has a uniLSTM with hidden size of 256 followed by a two dimensional linear layer. We jointly train the model for phonetic transcription and phrase discrimination task in MTL manner. While the phonetic training data remains the same as previous phase, we add 40,000 false trigger utterances and 140,000 true trigger utterances for the phrase discrimination task similar to \cite{Sigtia_2020}\cite{adya2020hybrid}\cite{dighe2020knowledge}\cite{sigtia2020progressive}.

\section{Evaluations and Ablation Study}
\label{sec:results}

\begin{figure}
\centering
\includegraphics[width=\linewidth, height=0.75\linewidth]{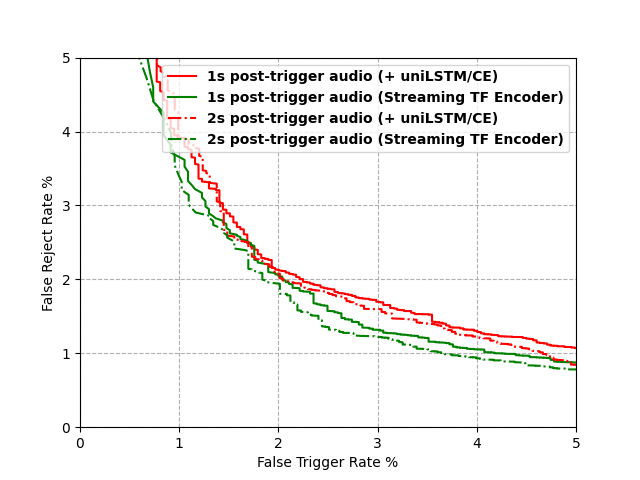}
\caption{False Trigger Mitigation with respect to post-trigger audio context.}
\vspace{-3mm}
\label{fig:det_curve_ftm}
\end{figure}

\begin{figure*}[htpb!]
\vspace{-3mm}
\centering
\begin{minipage}[b]{.48\textwidth}
\includegraphics[width=\linewidth, height=0.70\linewidth]{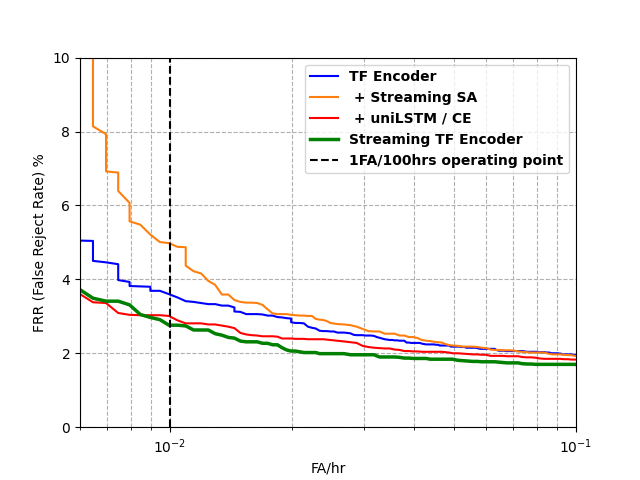}
\caption{DET curves for structured evaluation set}\label{fig:det_curve_checker_structured}
\end{minipage}\qquad
\begin{minipage}[b]{.48\textwidth}
\includegraphics[width=\linewidth, height=0.70\linewidth]{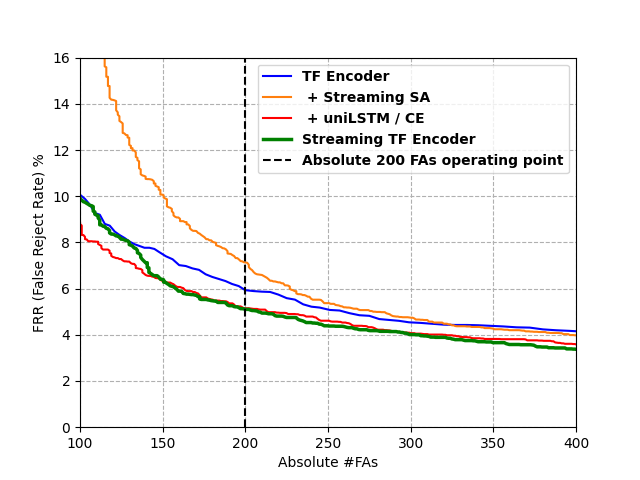}
\caption{DET curves for take-home evaluation set}\label{fig:det_curve_checker_unstructured}
\end{minipage}
\vspace{-3mm}
\end{figure*}

\vspace{1mm}
\begin{table*}[tpb!]
    \centering
    \begin{threeparttable}[b]
    \vspace{-3mm}
    \caption{Voice Trigger Detection performance for 4 models on 2 datasets \\ 
    }
    \small
    \centering
    \begin{tabular}{ p{0.25\textwidth} >{\centering}p{0.1\textwidth} >{\centering}p{0.12\textwidth} >{\centering}p{0.1\textwidth} >{\centering}p{0.12\textwidth} >{\centering\arraybackslash}p{0.12\textwidth} } 
    \hline
    Architecture                        & Streaming SA Layer    & uniLSTM in phrase discrimination  & Phrase discrimination loss   & FRR (\%) on structured dataset\tnote{1}  & FRR (\%) on unstructured dataset\tnote{2}\\
    \hline
    TF Encoder                      &               &               & CTC   & 3.6           & 5.9\\
    \  + \ Streaming SA             & \checkmark    &               & CTC   & 4.9           & 7.1 \\
    \  + \ uniLSTM \ / \ CE         &               & \checkmark    & CE    & 2.9           &5.1 \\
    \textbf{Streaming TF Encoder}   & \checkmark    & \checkmark    & CE    & \textbf{2.8}  & \textbf{5.1} \\
    \hline
    \end{tabular}
    \begin{tablenotes}
     \item[1]\footnotesize{FRR at an operating point of 1 FA/100 hrs for structured evaluation set }
     \item[2]\footnotesize{FRR at an operating point of 200 FAs on unstructured take home evaluation set}
    \end{tablenotes}
    \label{tab:vtd_frr}
    \end{threeparttable}
    \vspace{-5mm}
\end{table*}

\subsection{Evaluation Data}
Our proposed joint model is evaluated on 2 tasks: VTD and FTM. For VTD, we use the same evaluation datasets as used in our baseline system \cite{adya2020hybrid}. The first dataset is a structured evaluation set consisting of utterances from 100 participants recorded through a smart speaker with an equal proportion of male and female participants. Each of the utterance begins with the trigger phrase and is obtained under different noise and playback conditions. The negative data consists of 2,000 hours of general speech data which does not have the trigger phrase. The second unstructured dataset is recorded at home by 80 households. Each family uses a smart speaker daily for two weeks. A first-pass DNN-HMM system \cite{siri2017device} with a low threshold detects audio segments which are acoustically similar to the trigger phrase, including false triggers. With this dataset, we can measure FRR and false-trigger rates for realistic in-home scenarios similar to in the wild usage. More detailed descriptions on the two evaluation datasets can be found in \cite{Sigtia_2020}\cite{adya2020hybrid}. For the FTM task, we use 88,000 true triggers and 2,800 unintended false triggers as evaluation data.

\subsection{Results}

\subsubsection{Voice Trigger Detection}

We evaluate the phonetic transcription branch of four different variants of our model to better understand the impact of each change from the baseline TF encoder to the proposed streaming TF encoder. 
\begin{itemize}
  \item TF encoder
  \item \ +\ streaming SA: vanilla SA layers in the encoder are replaced with streaming SA layers (Section \ref{sec:streaming_sa})
  \item \ +\ uniLSTM / CE: uniLSTM layer is added in the phrase discrimination branch (Section \ref{sec:lstm_disc_branch}) and the CTC is loss replaced by frame-wise CE loss (Section \ref{sec:ce_loss_disc_branch})
  \item Streaming TF encoder
\end{itemize}

Figure \ref{fig:det_curve_checker_structured} shows the Detection Error Trade-off (DET) curves on the structured evaluation set with log(X)-axis representing false alarms (FA) per hour of active audio and Y-axis representing FRR. While changing the vanilla SA layers to streaming SA layers degrades FRR, the introduction of a uniLSTM and CE loss in the phrase discrimination branch improves VTD task accuracy. In fact, our proposed model performs at par with the the case when SA layers are not streaming.

This is an interesting result since structure of the phonetic transcription branch remains unchanged; uniLSTM and CE loss changes were introduced to the phrase discrimination branch. However, since these two branches share a common backbone (blue block in Figure \ref{fig:model_arch}), better training of the phrase discrimination due to uniLSTM/CE loss ensures VTD improvements for phonetic transcription branch. Table \ref{tab:vtd_frr} shows FRR of all four variants at an operating point of 1 FA in 100 hours of active audio.

Figure \ref{fig:det_curve_checker_unstructured} shows the DET curves for the unstructured evaluation set. We observe a similar trend as seen for the structured evaluation set. Table \ref{tab:vtd_frr} shows the FRR of all four variants at an operating point of 200 false alarms. We have reported results for a different operating point here compared to \cite{adya2020hybrid} since our unstructured evaluation set has increased in past year.

\subsubsection{False Trigger Mitigation}

Next, we present the results for the FTM task. 
The phrase discrimination branch of our proposed joint system acts on the post-trigger audio to mitigate the remaining false triggers. The decision score at a given frame is determined by averaging per-frame positive class scores of the previous 10 frames. Figure \ref{fig:det_curve_ftm} shows DET curves with the X-axis as false trigger rate and Y-axis as FRR. We can clearly see that post-trigger audio is significantly helping to reduce unintended false triggers. At an additional FRR of 1\%, our proposed model can mitigate up to 95.8\% of the unintended false triggers with an additional one second of post-trigger audio. Table \ref{tab:ftm_frr} shows how the false trigger rate improves as more post-trigger audio is provided. In fact, our proposed streaming TF encoder performs at par with the vanilla SA layers.

\vspace{1mm}
\begin{table}[tpb!]
\caption{False Trigger Mitigation results with respect to post-trigger audio. False Trigger Rate shown at FRR of 1\%}
\vspace{-3mm}
\small
\centering
\begin{tabular}{lcc}
\hline 
Architecture & 1s & 2s\\
\hline 
TF Encoder & 7.0 & 5.5\\
\ + \ Streaming SA & 7.7 & 7.7\\
\ + \ uniLSTM\ /\ CE & 5.5 & 4.7\\
\textbf{Streaming TF Encoder} & \textbf{4.2} & \textbf{3.7}\\
\hline
\end{tabular}
\vspace{-5mm}
\label{tab:ftm_frr}
\end{table}

\subsection{Hardware Efficiency}
In this section, we show that the proposed streaming TF encoder network is more efficient than a vanilla SA layer based TF encoder in terms of runtime memory usage and reusing the computations to have constant runtime complexity as input audio duration increases. A typical vanilla SA layer has a computational complexity of $O(T^2D)$ where T is the audio length and $D$ is the hidden dimension of SA layers. The proposed streaming architecture converts the quadratic relation with the audio length into linear relation $O(TDS)$ by introducing a constant block shift $S$. This is particularly useful when analyzing long segments, where the computation can now be distributed in time instead of having one big compute block. This also ensures constant compute requirements at well defined points and hence compute resources can be reserved and scheduled more intelligently. Latency is another aspect to consider when using this technology for FTM. The streaming nature of our proposed model allows us to take decisions at regular intervals and cancel requests as soon as the score drops below a threshold. We compare our proposed model against the case when SA layers process the full audio utterance at once. We use the same on-device experimental setup of a 2019 smart phone as our baseline \cite{adya2020hybrid}. Table \ref{tab:device_runtime} shows 56\% improvement in inference time when processing an additional one second of post-trigger audio. Network latency further improves by 63\% as we increase the post-trigger audio to two seconds. Network runtime memory usage also decreases by 32\%  with additional one second of post-trigger audio and 47\% with two seconds of post-trigger audio.

\vspace{1mm}
\begin{table}[tpb!]
\caption{Relative improvements in network runtime latency and memory usage on 2019 smart phone with respect to the additional post-trigger audio context. Our proposed streaming TF encoder is compared against the baseline TF encoder}
\vspace{-3mm}
\small
\centering
\begin{tabular}{lcc}
\hline 
Metric & 1s & 2s\\
\hline 
Network runtime memory & 32\% & 47\%\\
Network runtime latency & 56\% & 63\%\\
\hline
\end{tabular}
\vspace{-5mm}
\label{tab:device_runtime}
\end{table}

\section{Conclusions}

We have presented a unified and hardware-efficient streaming TF encoder architecture for VTD and FTM. We extended our baseline TF encoder by: (1) replacing vanilla SA layers in the encoder with streaming SA layers, (2) introducing frame-wise CE loss rather than the CTC loss in the phrase discrimination branch, and (3) adding a uniLSTM layer in the phrase discrimination branch. The proposed model improves FRR by an average 18\% relative and correctly mitigates up to 95\% of unintended false triggers with an additional one second of post-trigger audio. Additionally, our model also shares the compute between VTD and FTM tasks leading to 32\% reduction in on-device run time memory and 56\% reduction in inference time compared to the baseline.

\bibliographystyle{IEEEtran}

\bibliography{mybib}

\end{document}